\documentstyle[psfig,twocolumn,aps]{revtex}
\begin{document}

\draft
\title{Polaron and Bipolaron Defects in a Charge Density Wave:
a Model for Lightly Doped BaBiO$_3$}

\author{ Ilka B. Bischofs\cite{byline1}, 
Vladimir N. Kostur\cite{byline2}, Philip B. Allen}
\address{Department of Physics and Astronomy, State University of New York,
Stony Brook, New York 11794-3800}

\date{\today}

\maketitle

\begin{abstract}

BaBiO$_3$ is a prototype ``charge ordering system'' forming
interpenetrating sublattices with nominal valence Bi$^{3+}$
and Bi$^{5+}$.  It can also
be regarded as a three-dimensional version of
a Peierls insulator, the insulating gap being a consequence of an
ordered distortion of oxygen atoms.  When
holes are added to BaBiO$_3$ by doping, it remains insulating
until a very large hole concentration is reached, at which point
it becomes superconducting.  The
mechanism for insulating behavior of more lightly-doped samples is formation
of small polarons or bipolarons.  These are self-organized point defects
in the Peierls order parameter, which trap carriers in bound states
inside the Peierls gap.  
We calculate properties of the polarons and bipolarons
using the Rice-Sneddon model.  Bipolarons are the stable defect;
the missing pair of electrons come from an empty midgap state
built from the lower Peierls band. 
Each bipolaron distortion also pulls down six localized states below
the bottom of the unoccupied upper Peierls band.
The activation energy for bipolaron hopping is estimated.

\end{abstract}

%\narrowtext

\section{The Model}

Pure BaBiO$_3$ has a distorted perovskite structure \cite{Cox}. The nominal
valence of Bi is 4+, leaving a single $s$-electron per Bi atom.
If BaBiO$_3$ had perfect cubic perovskite symmetry, then
(in independent electron picture, or band theory) it would be
metallic, with the Fermi level half into a broad
band made of Bi $6s$ electrons 
(partially antibonded with O $2p_{\sigma}$ states.)
The large size of the Bi $6s$ orbital indicates that
single-electron (band) approximation should be good \cite{Vielsack}.  Therefore
the origin of insulating behavior should lie in the structural
distortions which double the unit cell.  The simplest interpretation
is that BaBiO$_3$ is a prototype d=3 Peierls insulator with a
simple lattice dimerization.

BaBiO$_3$ can be hole-doped by Pb substitution for Bi or K substitution
for Ba.  At a quite high critical concentration $x_c$ of holes
per cell ($x_c \approx 0.65$ for Pb or $\approx 0.35$ for K),
an insulator to metal transition occurs, with the metal a
superconductor.  The high superconducting transition temperature
($T_c \le $ 30K) is compatible with a conventional electron-phonon
mechanism, but neither theory \cite{Meregalli} nor spectroscopy 
\cite{Timusk} is able to make a convincing confirmation.
Electron-phonon effects certainly play an important
role in the insulating part of the phase diagram.  This paper 
explores the self-trapped polarons which occur at low doping and
which provide the most plausible mechanism explaining how these materials
remain insulating to such high hole concentrations $x_c$.

\begin{figure}[t]
\centerline{\psfig{figure=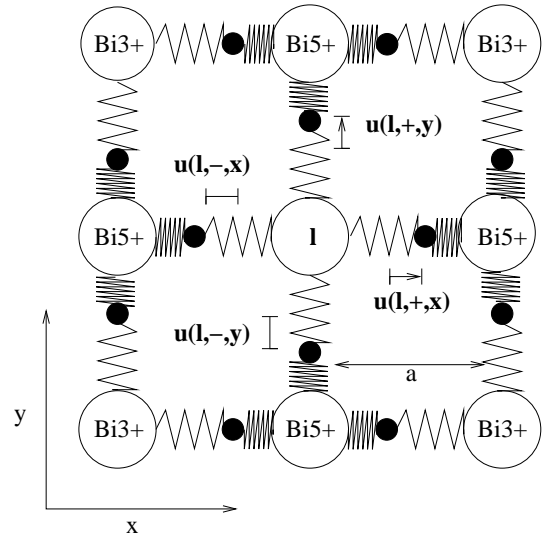,height=3.0in,width=3.0in,angle=0}}
\caption{Two-dimensional section of the Rice-Sneddon model.  The small
filled circles are oxygen atoms which are displaced by amounts $u$
from the midpoints of the bonds.  The central Bi atom (labeled $\vec{l}$)
is on the A sublattice (nominally Bi$^{3+}$) where electrons accumulate
because oxygens have moved away, lowering the potential energy of electrons.}
\label{fig:model}
\end{figure}

We use a simple and yet fairly realistic model \cite{Rice} which gives
a microscopic description of the Peierls distortion.  The model
is illustrated in Fig. \ref{fig:model}.
There is one electronic degree of
freedom per Bi atom, the amplitude of the $6s$ electron
orbital at that site, with corresponding creation operator
$c^{\dagger}_{\ell}$.  The index $\ell$ is a composite,
standing for $(\vec{\ell},\sigma)$ where the integer
vector $\vec{\ell}=(\ell_x,\ell_y,\ell_z)$ locates
Bi atoms on a cubic lattice of lattice constant $a$=4.28$\AA$,
and $\sigma=\pm 1$ is the spin index.  
The Bragg vector of the dimerized lattice is $\vec{Q}=(\pi,\pi,\pi)$.
The phase factor $\exp(i\vec{Q}\cdot\vec{\ell})=(-1)^L$, 
where $L=\ell_x+\ell_y+\ell_z$,
separates the simple cubic lattice into two sublattices, $A$ and $B$, 
with $(-1)^L$ equal to 1 on $A$ and -1 on $B$.  The 6 nearest
neighbors of each atom lie on the opposite sublattice.

\par
\par

In each cell there are three vibrational degrees of
freedom, the oxygen displacements along the Bi-O bond directions.
Vibrations perpendicular to bond directions are assumed not to influence
electrons.  The oxygen displacements $u(\vec{\ell},\alpha)$
are labelled by indices $\alpha=x,y,z$
which refer to the oxygen located at position
$a\vec{\ell}+(a/2)\hat{\alpha}$. 
It is convenient to define a local dilation or breathing
amplitude $e(\vec{\ell})$ on the $\vec{\ell}$-th Bi atom,
\begin{equation}
e(\vec{\ell})=\sum_{\alpha=x,y,z}[u(\vec{\ell},+,\alpha)-u(\vec{\ell},-,\alpha)]
\label{eq:dilation}
\end{equation}
where $u(\vec{\ell},\pm,\alpha)$ is the displacement of
the oxygen atom at $a\vec{\ell}\pm(a/2)\hat{\alpha}$.  Thus
$u(\vec{\ell},+,\alpha)$ is the same as  $u(\vec{\ell},\alpha)$, while
$u(\vec{\ell},-,\alpha)$ is the same as 
$u(\vec{\ell}-\hat{\alpha},\alpha)$,
that is, it is the displacement of the oxygen atom located
at $-(a/2)\hat{\alpha}$ relative to the
Bi atom $\vec{\ell}$.  For a positive distortion $e(\vec{\ell})$, the Bi $6s$
electron has its energy lowered by $-ge(\vec{\ell})$ where $g$
is the deformation potential.
Each oxygen displacement
costs elastic energy $(K/2) u(\vec{\ell},\alpha)^2$. 
Using the Raman measurement of $\hbar\omega$=70 meV for the oxygen
``breathing mode'' \cite{Raman}, the value of $K$ is 19 eV/$\AA^2$.
The resulting Hamiltonian is
\begin{equation}
{\cal H}=-t\sum_{<\ell,\ell'>}c^{\dagger}_{\ell} c^{}_{\ell'}
-g\sum_{\ell}e(\vec{\ell})c^{\dagger}_{\ell} c^{}_{\ell}
+\frac{1}{2} K \sum_{\vec{\ell},\alpha}u(\vec{\ell},\alpha)^2
\label{eq:ham}
\end{equation}
where the hopping summation index $<\ell,\ell'>$ goes over each nearest
neighbor pair both forward and backward, and conserves spin.
The variables $e(\vec{\ell})$ for different $\vec{\ell}$ are not independent,
only the original variables $u(\vec{\ell},\pm,\alpha)$ of Eq.
(\ref{eq:dilation}) are independent.  Piekarz and Konior \cite{Piekarz}
have introduced new decoupled breathing mode variables, at the 
expense of more distant terms in the coupling (second term) of
Eq. (\ref{eq:ham}).

For the time being we make the adiabatic approximation, treating
oxygen mass $M$ as infinite (neglecting oxygen kinetic energy.)
The electronic band structure is modelled 
with a simple $(ss\sigma)$ hopping integral $t$ between 
nearest neighbor Bi atoms.  When there is no oxygen displacement, this
yields the dispersion
\begin{equation}
\epsilon(\vec{k})= -2t(\cos(k_x)+\cos(k_y)+\cos(k_z))
\label{eq:epsilon}
\end{equation}
Comparison with band theory
\cite{Meregalli,Mattheiss,Liechtenstein,Blaha,Kunc} 
indicates that the value $t = 0.35 \pm 0.5$ eV
is appropiate for the band-width $12t \approx$ 4 eV of the conduction
band in cubic BaBiO$_3$.  Our values for the parameters are
summarized in table \ref{table:parameters}.

\begin{table}
\caption{Estimated values of parameters for BaBiO$_3$.
All parameters have $\approx$10\% uncertainty.}
\label{table:parameters}
\begin{tabular}{|rcl|}
%\hline
\multicolumn{3}{|c|} {Fitted Parameters}\\
$t$ & = & 0.35 eV \\
$K$ & = & 19 eV/$\AA^2$ \\
$g$ & = & 1.39 eV/\AA \\
\hline
\multicolumn{3}{|c|} {Derived Parameters}\\
$\Gamma=g^2/Kt$ & = & 0.30 \\
$\delta=\hbar\omega/t$ & = & 0.017 \\
%\hline
\end{tabular}
\end{table}

If energies are measured
in units of $t$ and lattice displacements in units $\sqrt{t/K}$,
then the model contains only a single dimensionless coupling constant
$\sqrt{\Gamma}$ where $\Gamma=g^2/Kt$.  
The dimensionless Hamiltonian ${\cal H}/t$
is just Eq.(\ref{eq:ham}) with the substitutions $t \rightarrow 1$,
$K \rightarrow 1$, and $g \rightarrow \sqrt{\Gamma}$.
Later when non-adiabatic
effects are treated, a new dimensionless energy ratio 
$\delta=\hbar\omega/t$ enters.

Because hopping goes only
between inequivalent sublattices, the band energy has the symmetry property
$\epsilon(\vec{k}+\vec{Q})=-\epsilon(\vec{k})$.
The bands are symmetric around energy 0, which is the Fermi
energy at half filling.  There is perfect ``nesting'' at this
value of the Fermi energy, because for every state with $\epsilon_k=0$,
there is another state at wavevector $\vec{k}+\vec{Q}$
which is also at the Fermi energy.  This perfect nesting property
is an artifact of the nearest neighbor hopping, and is destroyed
by next-neighbor hopping terms.  Because of perfect nesting,
it is particularly simple to find the ground state of the undoped
(half-filled) case, obtaining a BCS-like integral equation for
the Peierls gap $2\Delta$, which has a solution
$\Delta\approx 6.5t\exp(-0.291/\Gamma)$ for weak coupling ($\Gamma \ll 1$).
This was discussed in a previous paper \cite{Allen}.
The next section will treat this problem in the extreme atomic
limit ($\Gamma \gg 1$) where a particularly simple approximation
$\Delta=12\Gamma t$ applies.

An important consistency check on the Hamiltonian is the fact that
three parameters ($t$, $K$, and $g$) fit four independently 
determined properties of BaBiO$_3$.  The
value $g$=1.39 eV/$\AA$ is chosen to yield a Peierls band gap $2\Delta=$
2 eV, agreeing with optical measurements \cite{Opt}.  Then the
oxygen sublattice displacement $u_0=\Delta/6g$ is predicted by the model to
be 0.12 $\AA$, while experiment sees 0.09 $\AA$ \cite{Cox,Xray}.  

Our model Hamiltonian contains a good explanation of the difficulty 
of making free carriers: doped-in holes self-localize, making
small polarons and bipolarons.  In a future paper
we will re-examine the case of heavy doping, previously treated by
Yu, Chen, and Su \cite{Yu}.  Here we concentrate
on dilute doping.  We have previously
\cite{Kostur} looked at the criteria for polaron formation in
the case of dilute doping into an empty band (no Peierls distortion), 
as occurs in materials like BaSnO$_3$.  The critical
coupling strength $\Gamma_c(P)$ for polaron formation
was found to be 1.96, and $\Gamma_c(B)$ for bipolaron formation
was found to be 0.99.  We have also studied
the case of dilute doping into the half-filled band by an analytic
variational approximation \cite{Allen}.  Critical coupling strengths
were dramatically reduced, to $\Gamma_c(P)=0.18$ and
$\Gamma_c(B)=0.15$.  Here we provide detailed
numerical pictures of the behavior of lightly doped BaBiO$_3$,
including predictions about the optical spectrum.  In a companion
paper \cite{next} we examine the self-trapped exciton state which forms
when undoped BaBiO$_3$ is excited optically.

\section{Extreme Atomic Limit $\Gamma \gg 1$}

The value $\Gamma \sim 0.30$ is appropriate for BaBiO$_3$.  This is
intermediate between weak and strong coupling.  Beyond its
applicability to BaBiO$_3$, the model
is intellectually interesting in its own right, and we
can be curious how its behavior evolves with coupling
constant $\Gamma$ over the whole range 0 to $\infty$.
We now solve the problem in the extreme atomic 
limit $\Gamma \rightarrow \infty$
which just means that we ignore the hopping term ($t\rightarrow 0$),  
simplifying the mathematics greatly.
The results yield insight into the intermediate region
$\Gamma \sim 0.30$.

\subsection{Bipolaron crystal}

When the Bi $s$ level is half-filled, the strong-coupling solution
consists of putting two electrons on each Bi atom on the $A$ sublattice
and no electrons on each Bi atom on the $B$ sublattice.  This is the
extreme atomic limit of a Peierls charge density wave.  The paired electrons
can be given various names: ``bipolarons'' or ``negative $U$ centers.''
The mechanism stabilizing these objects is oxygen motion.
Each oxygen has as nearest neighbors one $A$ and one $B$ sublattice Bi atom.
Therefore it experiences a force away from the more negative $A$ site and 
toward
the more positive $B$ site.  The piece of the Hamiltonian (\ref{eq:ham})
containing this force is 
\begin{equation}
-gu(\vec{\ell},\alpha)\sum_{\sigma}(c^{\dagger}_{\ell} c^{}_{\ell}
-c^{\dagger}_{\ell^{\prime}} c^{}_{\ell^{\prime}})
\label{eq:force}
\end{equation}
where $\vec{\ell}^{\prime}=\vec{\ell}+\hat{\alpha}$ is the 
other member of the pair of nearest Bi atoms.  Let us define
the Peierls order parameter (amplitude of charge density wave) $\rho_0$ by
\begin{equation}
\sum_{\sigma}<c^{\dagger}_{\ell} c^{}_{\ell}>=1+(-1)^L \rho_0.
\label{eq:orderp}
\end{equation}
In the undistorted cubic structure, $\rho_0=0$.
In strong coupling, $\rho_0 \rightarrow 1$, that is, the Bi atoms have charge
2 on sublattice $A$ and 0 on sublattice $B$.  The force
is $2g\rho_0$ on each oxygen, directed
toward its nearest $B$-type Bi neighbor, counteracted by a
harmonic restoring force $Ku$, yielding an optimum displacement
$u_0 = 2g\rho_0/K$ or a ``breathing amplitude'' $e(\vec{\ell})=(-1)^L e_0$,
where, the oxygen breathing order parameter
$e_0$ is $12g\rho_0/K$.  These relations remain true for arbitrary coupling
($0 \le \rho_0 \le 1$).
In strong coupling, site energies of Bi $s$ orbitals are thus $-12g^2/K$
on $A$ sites and $+12g^2/K$ on $B$ sites.  For hopping
$t=0$, the lowest energy excitation is to move one electron
from an $A$ site to a $B$ site.  This costs energy $2\Delta_0
=24g^2/K=24\Gamma t$ which is the value of the Peierls gap.
We shall use the notation $\Delta_0$ to denote the value of
the gap in the atomic limit, and $\Delta$ for the value
for the actual coupling strength under consideration. 
The limiting answer $2\Delta_0=24g^2/K$ can also be found from 
the integral equation (23) of ref.
\cite{Allen}, by taking $\Delta \gg t$.

We get a measure of the value $\Gamma_c$ where the crossover
occurs between strong and weak coupling by comparing total
energies.  In the atomic limit, this is $-6g^2/K$ per atom
($-12g^2/K$ per electron of on-site orbital energy, but at
a cost of $+6g^2/K$ per atom of elastic energy.)  In the opposite
($g$=0) limit, the electrons gain on average $-2t$ of delocalization
energy by band formation in the undistorted crystal.  These energies
are equal at $\Gamma_c=1/3$. 

So far we are safe in the adiabatic approximation, since the
vibrational energy $\hbar\omega$ is
small compared with the Peierls gap in strong coupling.  However,
the true lowest energy excitation will cost less
than $2\Delta$ once we allow relaxation of the atoms.  Thus the 
optical spectrum will have non-adiabatic 
Franck-Condon character, with minimum
excitation energy corresponding to a ``self-trapped exciton.''
The companion paper \cite{next} pursues this topic.  The rest of this paper
is devoted to the case where
one or two holes are doped into the half-filled band.  We will consider
both the ground state and the new excitations induced by the doping. 

\par
\begin{figure}[t]
\centerline{\psfig{figure=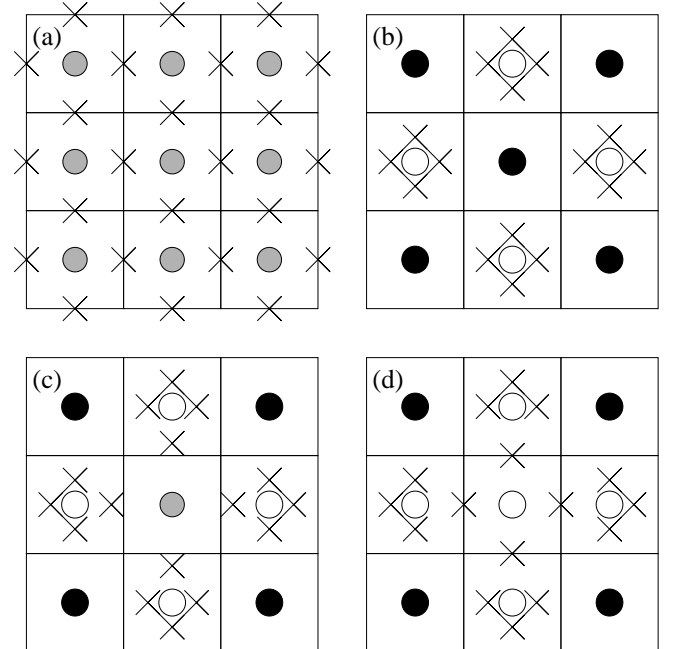,height=3.4in,width=3.37in,angle=0}}
\caption{Schematic structures: (a) is the undistorted high temperature
structure; (b) is the low temperature Peierls phase of pure BaBiO$_3$;
(c) has a hole polaron in the central cell; and (d) has a hole
bipolaron.  Circles denote Bi atoms, with filled circles being
Bi$^{3+}$ ions with two $s$ electrons, shaded circles being
Bi$^{4+}$ ions with one $s$ electron, and open circles being
Bi$^{5+}$ ions with no $s$ electrons.  X's denote oxygen ions.}
\label{fig:lattice}
\end{figure}
\par

\subsection{One hole in the atomic limit}

To remove an electron, keeping the Peierls distortion fixed,
costs energy $12g^2/K$ (the amount of Peierls attractive energy lost.)
When $t=0$, the hole sits on some particular
$A$ sublattice site as a singly charged Bi$^{4+}$ ion.
The 6 surrounding oxygens 
now experience a diminished repulsive force, $g$ instead of
$2g$, toward their Bi$^{5+}$ neighbors.  The local Peierls displacement
$u_1$ is therefore $g/K$, or half of $u_0$.  This reduces the elastic
energy by 6 times $K(u_0^2-u_1^2)/2$ or $9g^2/K$ but does this at the
expense of raising the energy of the $6s$ orbital at this site from
$-12g^2/K$ to $-6g^2/K$.  Since this orbital is occupied once, the
net energy saved by this local distortion is $3g^2/K$.  
The resulting object, a point defect in the Peierls order parameter,
is a small hole-polaron.  In a system with $N$ Bi atoms, the isolated
polaron corresponds to the system having $N-1$ electrons.
Its energy $E[P]\equiv E_{\rm gs}(N-1)
-E_{\rm gs}(N)$ is $9g^2/K$, smaller than
the removal energy $12g^2/K$ when no relaxation is permitted.
The difference is the polaron trapping energy, $E_t[P]=3g^2/K$.

Photoemission would measure the polaron formation energy $9g^2/K$ rather
than $12g^2/K$ provided the removal were done slowly enough that oxygen
atoms could move to their new optimum positions.  In fact, this is surely
not the case, and one expects to see instead a peak photoemission intensity
at the adiabatic energy $12g^2/K$, with a series of Franck-Condon vibrational
sidebands with a Gaussian envelope extending down to the threshold
$9g^2/K$.  A detailed discussion of the corresponding effect in LaMnO$_3$
was given earlier \cite{Perebeinos}.

\begin{figure}[t]
\centerline{\psfig{figure=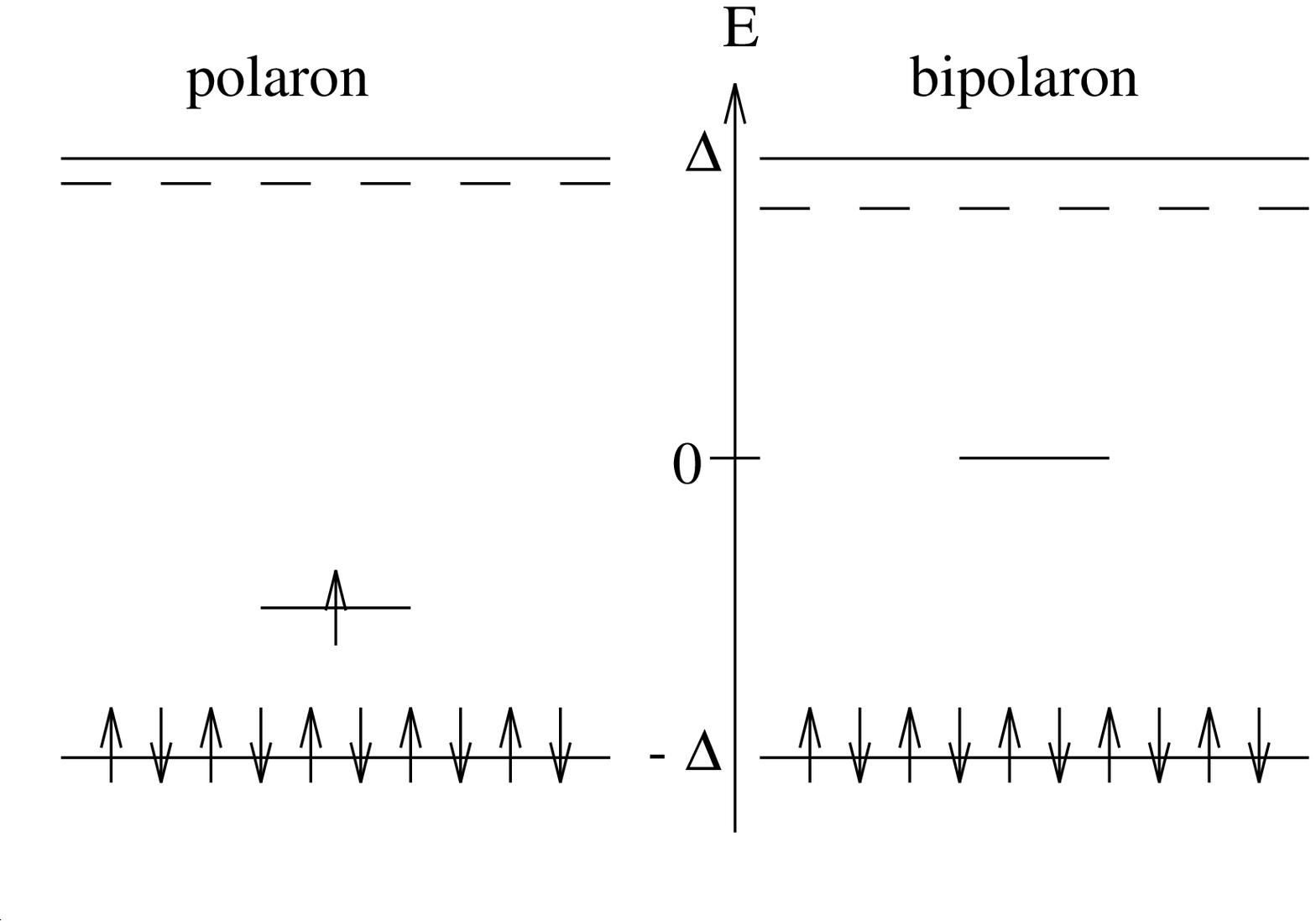,height=2.2in,width=3.0in,angle=0}}
\caption{Energy spectrum of the Rice-Sneddon model in the atomic
limit ($t=0$) with one hole present (left side labeled ``polaron'') or
with two holes present (right side labeled ``bipolaron.'')  In both
cases, the valence band contains $N/2-1$ doubly-filled states at
energy $-\Delta$, and
the conduction band has $N/2-6$ degenerate states at energy $\Delta$,
with 6 additional empty states bound below them.}
\label{fig:spectrum}
\end{figure}

The hole polaron is shown schematically in part (c) of Fig. 
\ref{fig:lattice}.  It is clear that the orbitals on the surrounding
6 Bi$^{5+}$ ions must also be altered by the oxygen relaxation.  Rather
than having their energy pushed up by $6g u_0$, their
energy is pushed up by $5g u_0  + g u_1$, or $11g^2/K$ rather than
$12g^2/K$.  Thus the polaron defect has an interesting spectrum of gap
states.  There is the singly occupied $A$-sublattice state 
located at $\Delta/2$ above the lower Peierls band at $-\Delta$,
and there are six empty $B$-sublattice states located at energy
$\Delta/12$ below the upper Peierls band at $\Delta$.
This spectrum is shown on the left side of Fig. \ref{fig:spectrum}.

\par
\par

\subsection{Two holes in the atomic limit}

Two polarons spatially separated cost two times $9g^2/K$, but if
they occupy the {\bf same} local site (with opposite spin, $S=0$) 
then it is easy
to see that the cost is only $12g^2/K$.  This is because removal
of two electrons costs $24g^2/K$ in lost Peierls energy, but
when sitting on a single $A$ site (converting the Bi$^{3+}$
to a  Bi$^{5+}$), the local strains on the six surrounding oxygens
all disappear ($u_2=0$.)  This gains back $12g^2/K$ that was previously
paid in strain energy.  The local $s$ orbital at the site of
the bipolaron now has energy 0 because of no local strain, but
unlike the polaron case, the loss of Peierls energy of this state
does not alter anything since the level is now empty.  Thus
the total energy of the bipolaron is $E[B]=12g^2/K$.  Relative
to free holes which cost $24g^2/K$, the
bipolaron trapping energy is $E_t[B]=12g^2/K$.  The energy released
when two polarons bind into a singlet bipolaron is $E_b=6g^2/K$. 
However, this
neglects on-site Coulomb repulsion $U$.  We expect a reduction of
the bipolaron binding energy to $E_b=6g^2/K - U$. 
The Hubbard $U$ is defined as the energy difference between charged
ions $E({\rm Bi}^{3+})+E({\rm Bi}^{5+})$ and ``neutrals''
$2E({\rm Bi}^{4+})$ for BaBiO$_3$ with oxygen atoms frozen in 
cubic perovskite positions.  
Experimentally, the absence of free spins in doped BaBiO$_3$
indicates that Bi$^{4+}$ (with $S=1/2$) is disfavored.  Therefore, $U$
cannot be large enough to destabilize
bipolarons.  Vielsack and Weber \cite{Vielsack} have done constrained density
functional calculations to obtain $U$.  They fit their results to an
effective $U$ of $0.6 \pm 0.4$ eV of a single band model like ours.
This is similar in size to our attractive energy of $6g^2/K \approx$ 0.6 eV.
When hopping is turned back on, the attractive energy goes down, but
the bipolaron also spreads out, so that the repulsive $U$ is
less effective.  Thus $U$ is small enough to permit bipolarons to exist.

The spectrum of BaBiO$_3$ with a bipolaron consists of a filled
lower Peierls band containing $N/2-1$ states at energy $-\Delta=-12g^2/K$,
one empty midband state $E=0$, and $N/2-6$ upper Peierls band
states at energy $+\Delta$.  Six new empty states appear
below the empty upper Peierls band, at energy $5\Delta/6$.  These
are the $s$ orbitals on the six $B$ sites surrounding the bipolaron,
which have only 5 out of 6 first neighbor oxygens with displacement 
$u_0$, and one undisplaced.  The spectrum is shown 
on the right side of Fig. \ref{fig:spectrum}.

\subsection{corrections for small hopping $t$}

For the perfectly ordered half-filled band,
the Peierls gap diminishes quadratically as $t=\Delta_0/12\Gamma$ is turned on,
\begin{equation}
\Delta = \Delta_0 \left( 1 - \frac{1}{48\Gamma^2} + \cdots \right).
\label{eq:gap}
\end{equation}
This follows by expanding the integral equation (23) of ref. \cite{Allen}
for small values of $t/\Delta$.  The amplitude $\rho$ of the CDW 
diminishes the same way, from 1 to $1-1/48\Gamma^2 +\cdots$.

The eigenstates are now extended Bloch states
of energy $\lambda(k)=\pm \sqrt{\Delta^2+\epsilon(k)^2}$, where
$\epsilon(k)$ is the eigenvalue of the undistorted lattice,
$-2t\sum_{\alpha}\cos(k_{\alpha})$.  Thus the
least expensive delocalized hole states are at the top of the lower Peierls 
band where $\epsilon(k)=0$ and cost $(12g^2/K)(1-1/48\Gamma^2)$.
The competing bipolaron state can no longer be completely localized
on a site because the charge on the nominal Bi$^{3+}$ ions
is $1+\rho <2.$  It is not easy to develop a systematic $1/\Gamma$
expansion for the localized solutions.  The problem is that
as oxygens move and the polaron localizes,
wavefunctions and charges of all occupied delocalized states must
be altered self-consistently near the polaron.

\section{numerical study of polaron}
\label{sec:numerical}

The algorithms used are explained in detail in the MA thesis of
the first author \cite{Bischofs}.  A large supercell (typically
between 100 and 1000 Bi atoms) was chosen, with periodic
boundary conditions applied.  It was useful to have the
translation vectors $\vec{A},\vec{B},\vec{C}$ of the supercell
{\bf not} lie along symmetry axes, but to use small asymmetrical deviations
in order to lift degeneracies.  The translation
vectors all translate $A$ sublattice points to other $A$
sublattice points, so that the Peierls state fits the cell.
Trial oxygen positions are chosen, the Hamiltonian matrix
is diagonalized, and forces on oxygen atoms are calculated
by summing over occupied wavefunctions to get the electronic
charges on the two adjacent Bi atoms.  A variable metric method \cite{Num}
was used to iterate towards optimal oxygen displacements.
We confirmed that for the half-filled case, the density of states,
oxygen coordinates, and Peierls gap all agreed well with calculations
by other methods.  The half-filled ground state serves as the
``vacuum'' for the rest of our work.
To study doping by one or two holes per
supercell, a cell with $\approx$200 atoms is sufficient if the coupling
constant $\Gamma>0.22$.  For smaller $\Gamma$, larger cells
are needed.

\par
\begin{figure}[t]
\centerline{\psfig{figure=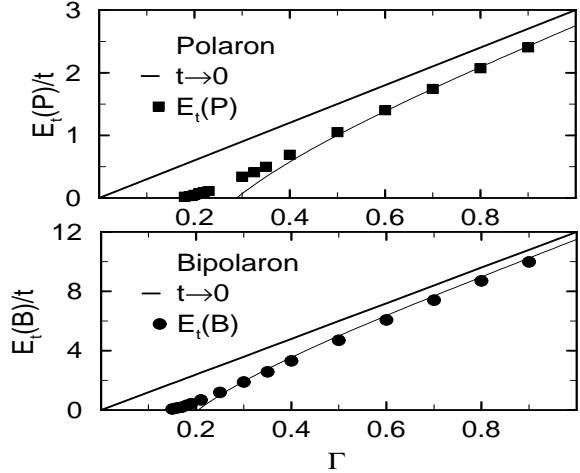,height=2.6in,width=3.6in,angle=0}}
\caption{Trapping energy of a polaron (upper panel) and a bipolaron
(lower panel) computed numerically.  Bold solid lines are strong-coupling
predictions.  Thin lines are approximate $1/\Gamma^2$ corrections
discussed in Sec. III.A.}
\label{fig:polaron}
\end{figure}
\par

\subsection{total energy}

By simply restricting the occupancy to be $N-1$ or $N-2$, we obtain
solutions with polaron or bipolaron local defects.
The numerical results for the ``trapping energies''
are shown in Fig. \ref{fig:polaron}.
Stable polaron solutions occur when $\Gamma  > \Gamma_c(P)\approx 0.18$.
Stable bipolaron solutions occur when $\Gamma  > \Gamma_c(B)\approx 0.15$.
At lower values of $\Gamma$ an inserted hole just stays at the top
of the valence band, causing an order $1/N$ renormalization of the 
Peierls gap.  For $\Gamma  > \Gamma_c$, such states were always
found to be numerically unstable toward formation of local defects.
The onset of localized solutions seems to occur continuously, 
unlike the case where an empty $s$-band is doped and localized
solutions appear discontinuously (in adiabatic approximation)
at much larger values $\Gamma_c = 1.96$.
There are several reasons for the ease of polaron formation 
in the Peierls state relative to the empty band.
Perhaps foremost is the fact that there is less to be gained
by delocalization, because low-energy delocalized states are
occupied.  Also, the Peierls state has
lattice strain already built in, so polaron formation can occur
by reduction of the Peierls lattice strain 
rather than by initiation of new lattice strain.

Although we do not have a systematic $1/\Gamma$ perturbation theory,
nevertheless, a very simple correction to the $\Gamma\rightarrow\infty$
expression can be made.  The trapping energy of polarons 
$E_t[P]$ is the difference between the energy $\Delta$ to create
a hole without relaxation, and the energy $E[P]$ of the relaxed
polaron.  In the $\Gamma\rightarrow\infty$ limit, this
gives $E_t[P]/t=12\Gamma - 9\Gamma$.  The first term, $\Delta_0/t$,
the energy to put a hole into the ``vacuum'' without
lattice relaxation,  can be corrected by Eq. (\ref{eq:gap}) to
$\Delta(\Gamma)/t \approx 12\Gamma - 1/4\Gamma$.  We do not know the form
of the correction to $E[P]$, but find that the correction is weaker
than the vacuum correction.  Therefore, correcting only the vacuum
term gives the formula $E_t[P]/t \approx 3\Gamma - 1/4\Gamma$.
The $\Gamma\rightarrow\infty$ formula (labeled $t\rightarrow 0$)
and the vacuum-corrected formula are shown in Fig. \ref{fig:polaron}
as thick and thin solid lines.  The fit to the polaron is very good
down to $\Gamma=0.5$.  For the bipolaron, the corresponding 
formulas are $E_t[B] \equiv 2\Delta-E[B]$, and 
$E_t[P]/t \approx 12\Gamma - 1/2\Gamma$.  The vacuum-corrected fit
is not as good as for the polaron.

\subsection{size and shape of polarons}

\begin{figure}[t]
\centerline{\psfig{figure=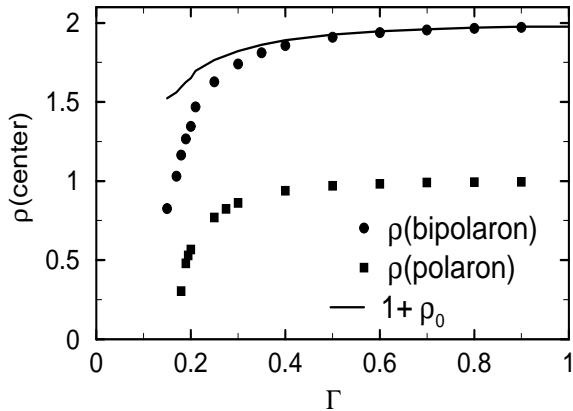,height=2.4in,width=3.3in,angle=0}}
\caption{The total charge removed from the central site of the 
hole polaron and bipolaron.  The solid line is the total charge
at that site in the Peierls state, the maximum available for removal.}
\label{fig:centralch}
\end{figure}

In the $t\rightarrow 0$ limit, an inserted hole sits
on a single site.  Hopping causes the hole to 
spread, gaining delocalization energy.
The amount of positive hole charge $\rho(\vec{r})$ 
at a site $\vec{r}$ is defined as
\begin{equation}
\rho(\vec{r})=\sum_{i}^{\rm{occ, vac}} |\Psi_{i,{\rm vac}}(\vec{r})|^2
-\sum_i^{\rm{occ},P} |\Psi_{i,P}(\vec{r})|^2,
\label{eq:rhocenter}
\end{equation}
where ``vac'' and $P$ indicate the single particle states 
$\Psi$ for vacuum and polaron cases.  In Fig. \ref{fig:centralch},
the hole charge at the central site $\vec{r}=0$ is plotted.
For $\Gamma>0.5$,
the polaron is almost completely on the central site,
whereas for the bipolaron, the available charge on the central
$A$-type atom is essentially all depleted for $\Gamma>0.4$,
but until a somewhat larger $\Gamma$, some charge must
be removed from surrounding sites.  At the realistic
value $\Gamma=0.3$, the polaron depletes 0.84 electrons
from the central A site, and the bipolaron depletes 1.74
electrons, which is almost all of the 1.80 electrons which
occupy A sites in the vacuum state.

\par
\par

\begin{figure}[t]
\centerline{\psfig{figure=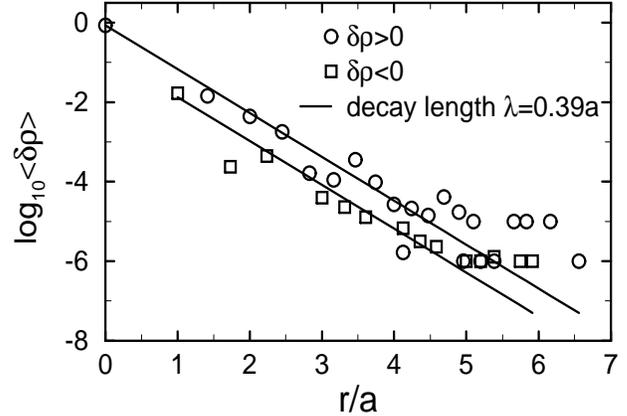,height=2.4in,width=3.3in,angle=0}}
\caption{Charge versus distance for the hole polaron when $\Gamma=0.30$.
The calculation used a cell with $N$=1000.
Sites with depletion of electron density (charge $\delta\rho>0$) are denoted
by circles; sites with excess electron density ($\delta\rho<0$)
are denoted by squares.  Beyond $r=5a$, overlapping charge from
neighboring cells affects the answer.}
\label{fig:chdecay}
\end{figure}

The polaron defect diminishes the amplitude
of the Peierls charge-density wave in its local vicinity.
This is seen by examining the charge on atoms near the polaron.
At $\Gamma=0.3$, to make a polaron, 0.16 electrons are taken from 
sites other than the central site.  
We find that neighboring B
sites do {\bf not} lose electron density, but rather gain electrons
when a polaron is formed nearby; neighboring A sites
are more depleted of electrons than if B sites had been unaffected.
The distance-dependence of the hole charge is shown in 
Fig. (\ref{fig:chdecay}) to be exponential, with decay length 0.39$a$.
The corresponding decay length for the bipolaron 
disturbance is 0.32$a$.  

\par
%\par

If small polaron formation occurs in
continuous fashion for $\Gamma > \Gamma_c$, as suggested by 
Fig. \ref{fig:centralch} shows the continuous evolution with
$\Gamma$, for $\Gamma > \Gamma_c$, from large CDW-like
(bi)polarons to small (bi)polarons.  The
decay length $\lambda$ (defined as the negative reciprocal
of $d\ln(\delta\rho)/dr$) should diverge as
$\Gamma$ is reduced to $\Gamma_c$.  This cannot be verified
numerically because of the small linear size of available
computational cells in 3 dimensions.  We do see a trend
in the right direction, shown in Table \ref{table:decay}.

\begin{table}
\caption{Decay constants for the hole charges for a (bi)polaron}
\label{table:decay}
\begin{tabular}{|ccccc|}
%\hline
$\Gamma$        &       $\lambda_p/a$ & $\rho_{\rm center}^p$ &
$\lambda_b/a$ & $\rho_{\rm center}^b$ \\
%\hline
%2              &       0.13    &       1       & &  \\
%\hline
0.9             &       0.21    &       0.99             &0.20  &1.97  \\
%\hline
0.6             &       0.24    &       0.98             &0.23  &1.94  \\
%\hline
0.3             &       0.39    &       0.84             &0.32  &1.74 \\
%\hline
0.2             &       0.57    &       0.58             &0.41  &1.40  \\
%\hline
\end{tabular}
\end{table}

\subsection{Conduction band edge states}

\begin{figure}[t]
\centerline{\psfig{figure=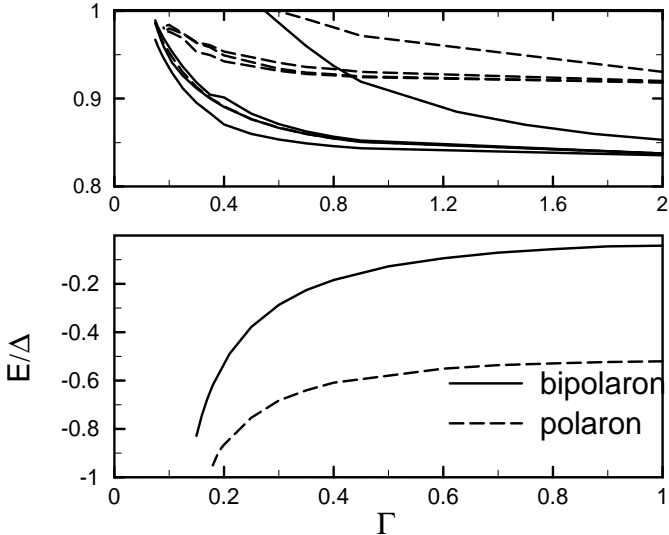,height=2.9in,width=3.6in,angle=0}}
\caption{Energies of localized single particle states lying in the gap.
The lower panel shows the states which derive from the valence band;
the polaron state is occupied once (spin degeneracy = 2.)  All other
states are empty in the ground state.  The upper panel, showing states
localized below the conduction band edge, has a different scale,
both horizontally and vertically.  }
\label{fig:gapstates}
\end{figure}

The polaron defect perturbs the band Peierls states in a local region.
This perturbation is sufficiently strong to create new localized states
at the bottom of the conduction band.  This effect is particularly 
simple in the strong-coupling limit, where we already discussed
(see Fig. \ref{fig:spectrum} and nearby text) six states bound by
$\Delta/12$ (near a polaron) or $\Delta/6$ (near a bipolaron).
These states can be classified as one $s$-like ($A_{1g}$), a
$d$-like doublet ($E_g$) and a $p$-like triplet ($T_{1u}$).
Therefore, as hopping $t$ is turned on, we can expect the six
bound states to split into singlet, doublet, and triplet, and
for weak coupling, perhaps to become unbound.  The numerical
results are shown in Fig. \ref{fig:gapstates}.  The $s$-like state
is bound only for $\Gamma > \Gamma_c(s) \approx$ 0.30 (for
a polaron) or 0.28 (for a bipolaron), while the
other states are nearly degenerate and look to persist all the
way to the critical $\Gamma$ for polaron or bipolaron formation.
Numerical finite-size effects prevent high accuracy in these
estimates.  The reason for the weaker binding of the $s$-like
state is that it couples to the polaron state which splits
off from the valence band, forming a spread out polaron with no
radial node, and an empty state bound below the conduction
band with a radial node.

\par
\par

In principle, these bound states can be seen spectroscopically,
as weak mid-gap absorption proportional to doping concentration.
We believe that bipolarons are the stable point defects.  At
$\Gamma$=0.3, the doped-in holes are bound at $\approx 0.7\Delta$
or 0.7 eV above the top of the lower Peierls band.  The optical
matrix element for exciting an electron from the valence
band into this state (alternately, for exciting the mid-gap
bound hole to the valence band) is zero in first order at the
$\vec{k}=(\pi/2,\pi/2,\pi/2)$ band edge.  There is a non-zero
matrix element to the $p$-like states split off from the bottom
of the upper Peierls band, with energy $\approx 1.88\Delta$ or
1.88 eV.  These numbers should not be taken as reliable predictions
of actual spectral lines for several reasons.  First, the predictions
are sure to be somewhat sensitive to the model, especially to
the fact that our model omits the disorder potential caused by
dopant atoms.  Second, there will be significant Franck-Condon
effects coming from the fact that if the empty midgap states are
suddenly occupied by an optical transition, local oxygen relaxations
can lower the energy of these states.  Finally, as explained in
the companion paper, the Peierls gap of pure BaBiO$_3$
in this model is not ``clean.''  When an electron-hole pair is created
in the pure material, self-trapping of the electron and hole should
occur, causing absorption in Franck-Condon sidebands below the energy
of the nominal Peierls gap $2\Delta$.  This may obscure additional
spectral features introduced by bipolarons formed by doping.

\section{Mobility of Bipolarons}

At $T$=0 K, band formation occurs, in principle, with
band width exponentially reduced by Huang-Rhys factors \cite{Huang,Mahan}
$\exp(-S)$ coming from vibrational overlap integrals.  Let us make
a crude estimate for the polaron.  To simplify, we work
in the strong coupling limit where the polaron charge is
all at one site, $\vec{\ell}$.  The state $|\vec{\ell}P>$
representing the polaron at this site, is a product of
a hole wavefunction $c_{\ell}|{\rm vac}>$ (where $|{\rm vac}>$
is the electronic vacuum) times a vibrational wavefunction
$|e_1(\vec{\ell})>$ (the vibrational ground
state with six oxygens in altered breathing positions around
site $\vec{\ell}$.)  We want the effective hopping matrix element
from this state to a degenerate polaron state $|\vec{\ell}^{\prime}P>$
located at a neighbor site $\vec{\ell}^{\prime}$.  The
Hamiltonian matrix element $<\vec{\ell}P| {\cal H} |\vec{\ell}^{\prime}P>$
factorizes into
\begin{equation}
t_{\rm eff}=<{\rm vac}|c_{\ell}^{\dagger}{\cal H}c_{\ell^{\prime}}|{\rm vac}>
	<e_1(\vec{\ell})|e_1(\vec{\ell}^{\prime})>
\label{eq:effhop}
\end{equation}

The first factor of Eq.\ (\ref{eq:effhop}) is the matrix element
to hop from $\vec{\ell}$ to $\vec{\ell}^{\prime}$, both
being on the A sublattice (since that is where the polarons reside.)
This is a second-neighbor hop on the underlying simple cubic lattice.
Our Hamiltonian omitted second-neighbor hops, but can now be
extended to include such a hopping matrix element $t^{\prime}$,
smaller than the first-neighbor hopping $t$.  Alternately, we could
do second-order degenerate perturbation theory, which would be
rather messy and would probably yield something of the same magnitude.

The second factor of Eq.\ (\ref{eq:effhop}) is the ``Huang-Rhys
factor,'' namely the overlap between vibrational ground states of
two different lattice configurations.  Twelve oxygen atoms change
locations between the two states.  Six oxygens around the original
polaron site change positions from $u_1$ to $u_0$, and six oxygens
around the new polaron site change positions from $u_0$ to $u_1$.
The corresponding overlap parameter is 
\begin{equation}
\exp(-S)=\exp[-12(M\omega/\hbar)(u_0-u_1)^2/4].
\label{eq:huang}
\end{equation}
The exponent, rewritten as $S=(\Delta/\hbar\omega)(\Delta/t)(1/48\Gamma)$,
is approximately 2.8.
The polaron band width is thus of order 0.06 eV,
if we use the band-width 4 eV, and assume that the second-neighbor
hopping is approximately $t^{\prime}=t/4$.  The disorder caused by the dopant
atoms is undoubtedly larger than this, so we should expect polarons
to be Anderson-localized.  

The bipolaron band-width will be much narrower.  For two electrons
to hop to second neighbors is a higher order process, and the Huang-Rhys
exponent $S$ is larger by 4 since oxygens move the full amount $u_0$
rather than by $|u_0-u_1|=u_0/2$.  Thus the Huang-Rhys factor for
bipolarons is smaller by $2\times 10^{-4}$ than for polarons, and
we expect bipolarons to be completely immobile at $T$=0 K.  

\begin{figure}[t]
\centerline{\psfig{figure=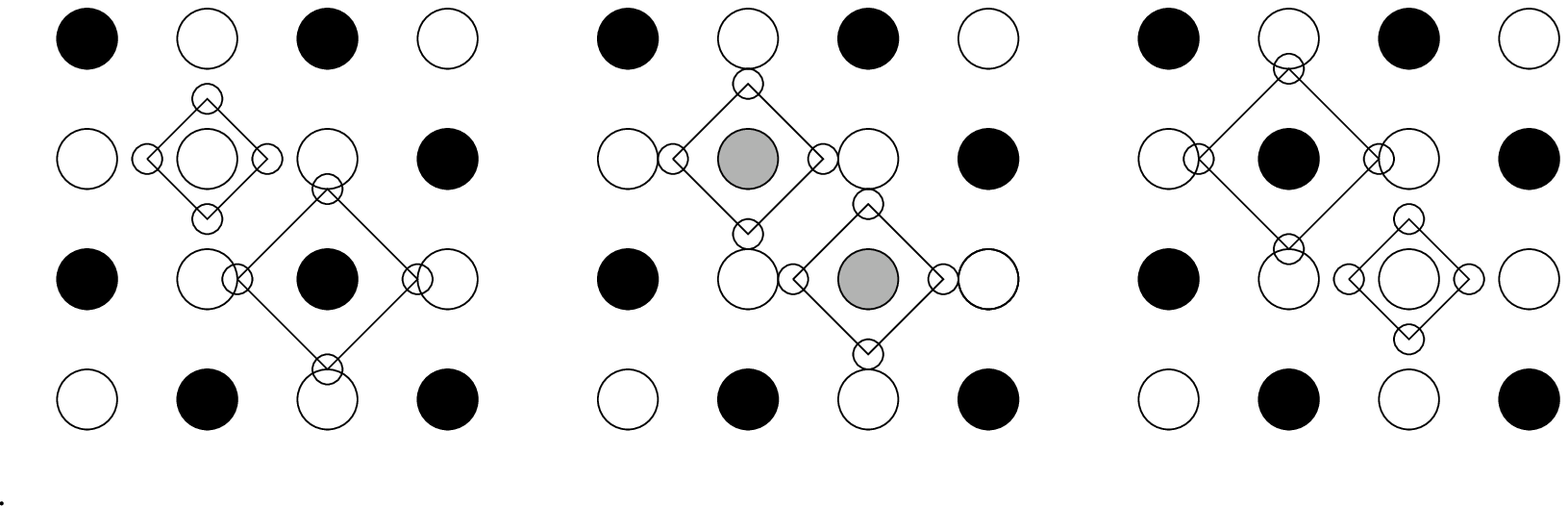,height=1.1in,width=3.47in,angle=0}}
\caption{Hopping of a small bipolaron through an intermediate
state with two polarons.  Filled circles are Bi$^{3+}$ ions with
two electrons, and open circles are Bi$^{5+}$ ions with
no electrons.  The left panel has a hole bipolaron in the small
square, which evolves into two hole polarons in the middle panel
(shaded circles, medium size squares.)  One of the polarons
then collapses back to a Bi$^{3+}$ ion (right panel, large square)
and the bipolaron has moved to a new location (small square).}
\label{fig:hopping}
\end{figure}

At high temperature ($k_B T \approx \hbar\omega$), bipolaron
mobility should reduce to the classical version, attempt
frequency times thermal activation factor.  Assuming that
thermal activation is given by $\exp(-E_S/k_B T)$, with
$E_S$ the energy of the saddle point, we can make a plausible
argument for how mobility should behave at high $T$.
The saddle point is illustrated in Fig. \ref{fig:hopping}.
In this state, the bipolaron has split into two polarons
located at A-sublattice first neighbors, with full lattice
relaxation to minimize the energy at the saddle.  This state
lies above the bipolaron state in energy by 
the energy required to unbind a bipolaron
into a first-neighbor polaron pair, $\Delta/2\approx$ 0.5 eV
in strong-coupling approximation.  For the actual value $\Gamma$=0.3,
we estimate the activation energy to be $E_S \approx \Delta/3 \approx$ 0.3 eV.
Experimental measurements \cite{resis} vary from 0.17 eV to 
0.27 eV.

\par
\par

\acknowledgements
We thank R. Bhargava for writing computer programs,
and V. Perebeinos and A. Abanov for help and encouragement.
This work was supported by NSF grant no. DMR-0089492.

%\begin{thebibliography}{99}

\end{document}